\documentclass[pra,twocolumn,showpacs]{revtex4}

\usepackage{graphicx}
\usepackage{amsmath}
\usepackage{times}
\usepackage{dcolumn}

\begin{document}

\title[Bogoliubov spectrum and Bragg spectroscopy]{Bogoliubov spectrum 
and Bragg spectroscopy \\ of elongated Bose-Einstein condensates}

\author{C. Tozzo}
 \email{tozzo@science.unitn.it}
\affiliation{Dipartimento di Fisica, Universit\`a di Trento, and 
Istituto Nazionale \\ per la Fisica della Materia, BEC-INFM Trento,
I-38050 Povo, Italy }

\author{F. Dalfovo}
\affiliation{Dipartimento di Matematica e Fisica, Universit\`a
Cattolica, Via Musei 41, 25121 Brescia and \\
Istituto Nazionale per la Fisica della Materia, Unit\`a di Brescia
and BEC-INFM Trento}

\date{\today}

\begin{abstract}

The behavior of the momentum transferred to a trapped Bose-Einstein 
condensate by a two-photon Bragg pulse reflects the structure of the 
underlying Bogoliubov spectrum. In elongated condensates, axial phonons 
with different number of radial nodes give rise to a multibranch
spectrum which can be resolved in Bragg spectroscopy, as shown by 
Steinhauer {\it et al.} [Phys. Rev. Lett. {\bf 90}, 060404 (2003)]. 
Here we present a detailed theoretical analysis of this process. We 
calculate the momentum transferred by numerically solving the 
time dependent Gross-Pitaevskii equation. In the case of a cylindrical
condensate, we compare the results with those obtained by linearizing 
the Gross-Pitaevskii equation and using a quasiparticle projection 
method. This analysis shows how the axial-phonon branches affect the 
momentum transfer, in agreement with our previous interpretation 
of the observed data. We also discuss the applicability of this type 
of spectroscopy to typical available condensates, as well as the 
role of nonlinear effects.  

\end{abstract}

\pacs{03.75 Fi}

\maketitle

\section{Introduction}
\label{sec:intro}

In a recent paper \cite{steinhauer2003} we showed that interesting 
features emerge in the momentum transferred to an elongated Bose-Einstein 
condensate by a two-photon Bragg pulse when the duration of the pulse
is long enough. On the basis of numerical simulations with the 
Gross-Pitaevskii (GP) equation, we interpreted those features as due to 
the structure of  the underlying Bogoliubov spectrum: the Bragg pulse can
excite axial phonons with different number of radial nodes, each one 
having its dispersion law. Since the energy difference between these 
branches is of the order of the radial trapping frequency, they can be 
resolved only by Bragg pulses with duration comparable with the radial 
trapping period, as observed in \cite{steinhauer2003}. In this work,
we present a more detailed theoretical analysis that supports
our previous interpretation and gives further information about the 
role played by Bogoliubov excitations in Bragg spectroscopy. 

A useful insight into this problem is obtained by studying the case of 
an infinite condensate, unbound along $z$ and harmonically trapped in 
the radial direction $\rho$. We show that the response of such a 
cylindrical condensate retains all the relevant properties needed to 
interpret the observed behavior of a finite elongated condensate. In 
addition to the numerical solution of the full time dependent GP 
equation, we explicitly determine the time evolution of the order 
parameter in the linear (small amplitude) limit, using the quasiparticle
projection method of Refs.~\cite{morgan,bbg}. The cylindrical geometry 
allows us to simplify the calculations and, more important, to make 
the connection between the momentum transferred and the dynamic structure 
factor more direct than for a finite condensate.  By comparing the 
predictions of this approach to the results of the numerical integration 
of the time dependent GP equation, one can also distinguish the different 
effects of linear and nonlinear dynamics. This analysis of cylindrical 
condensates is presented in the first sections of this paper, while 
section V is devoted to the behavior of finite elongated 
condensates, as the ones in the experiments of 
Refs.~\cite{steinhauer2003,mit,rehovot}.

\section{Bogoliubov spectrum of a cylindrical condensate}
\label{section1}

First we investigate the excitation spectrum of a cylindrical 
condensate. We use the mean-field Gross-Pitaevskii theory at zero
temperature, assuming the condensate to be dilute and cold enough 
to neglect thermal and beyond mean-field contributions. The starting 
point is the time dependent GP equation for the order parameter 
$\Psi({\bf r},t)$ of a condensate of $N$ bosonic atoms subject to
an external potential $V({\bf r})$ \cite{rmp}:  
\begin{equation}
i \hbar \frac{ \partial }{ \partial t} \Psi =
 \left( - \frac{ \hbar^2 \nabla^2 }{ 2m } + V
+g |\Psi|^2 \right) \Psi  \; ,
\label{eq:TDGP}
\end{equation}
where $g=4 \pi \hbar^2 a/m$, and $a$ is the $s$-wave scattering 
length, that we assume to be positive. The order parameter is 
normalized according to $\int d{\bf r} |\Psi|^2 = N$. 
What is usually named Bogoliubov spectrum is the result of an 
expansion of $\Psi$ in terms of a quasiparticles basis in the form
\cite{bogoliubov,pitaevskii,fetter}:
\begin{eqnarray}
\Psi({\bf r},t)& = &  e^{-i\mu t/\hbar} \{ \psi_0 ({\bf r}) + 
 \nonumber \\
 & \sum_j &  [ c_j  u_j({\bf r}) e^{-i \omega_j t} 
             + c_j^* v_j^*({\bf r}) e^{i \omega_j t} ] \} 
\; ,
\label{eq:linearized}
\end{eqnarray}
where the $c_j$'s are constants and $\mu$ is the chemical potential. 
The quasiparticle amplitudes $u$ and $v$ obey the following 
orthogonality and symmetry relations: 
\begin{equation}
\int\! d{\bf r} \{ u_i u_j^* - v_i v_j^* \} =  \delta_{ij} 
\; \; ; \; \; 
\int\! d{\bf r} \{ u_i v_j - v_i u_j \} = 0 \; . 
\label{eq:ortho}
\end{equation}
This expansion can be inserted into Eq.~(\ref{eq:TDGP}). At zero 
order in  $u$ and $v$, one gets the stationary GP equation:
\begin{equation}
\left( - \frac{ \hbar^2\nabla^2}{2m } + V({\bf r})  +
g |\psi_0({\bf r})|^2 \right) \psi_0({\bf r}) = \mu  
\psi_0({\bf r}) \; .
\label{eq:GP}
\end{equation}
Among the solutions of this equation one finds the ground state 
order parameter, which can be chosen real and equal to the 
square root of the particle density \cite{note-gs}.  The next 
order gives the coupled equations
\begin{eqnarray}
\hbar \omega_i u_i &=& [ H_0 - \mu + 2 g \psi_0^2 ]
u_i  + g  \psi_0^2 v_i 
\label{eq:coupled1}
\\
- \hbar \omega_i v_i &=& [ H_0 - \mu + 2 g \psi_0^2 ]
v_i + g  \psi_0^2 u_i  \; ,
\label{eq:coupled2}
\end{eqnarray}
where $H_0= - (\hbar^2/2m) \nabla^2 +  V({\bf r})$.
The solutions of these equations provide the spectrum of the excited 
states in the linear (small amplitude) regime. 

Now we treat the specific case of a cylindrical condensate, which is 
unbound along $z$ and harmonically trapped along the radial direction, 
$\rho = [x^2 + y^2]^{1/2}$. 
Let us write the trapping potential as $V({\bf r}) 
= V_T(\rho)=(1/2) m \omega_\rho^2 \rho^2$. The ground state order 
parameter is uniform along $z$, while the stationary GP equation for the 
radial part becomes
\begin{equation}
\left[-\frac{\hbar^2 \nabla^2_\rho}{2m} +V_T(\rho)+\frac{gN}{L}
\phi_0^2(\rho)\right] \phi_0(\rho)=\mu\phi_0(\rho)
\label{eq:phi0}
\end{equation}
where $N/L$ is the number of bosons per unit length and the function 
$\phi_0$ is chosen to be real and subject to the normalization condition  
$2\pi \int_0^{\infty}\! d\rho \,  \rho  \phi_0^2 =1$. In this geometry,
the quasiparticle amplitudes $u$ and $v$ can also be factorized. We are 
actually interested in axially symmetric states (azimuthal angular 
momentum equal to zero), since they are the only ones excited by a 
Bragg process in which the momentum is imparted along the $z$-axis.  
These states, characterized by the axial wavevector, $k$, and the number 
of radial nodes, $n$, can be written in terms of new functions
$u_{n,k}(\rho)$ and $v_{n,k}(\rho)$ in the form
\begin{equation}
(u,v)_{n,k}(\rho,z)= L^{-1/2} e^{ikz} (u,v)_{n,k}(\rho) \; . 
\label{eq:uv-cylinder}
\end{equation}
The Bogoliubov equations (\ref{eq:coupled1})-(\ref{eq:coupled2}) then 
become 
\begin{widetext}
\begin{eqnarray}
\hbar \omega_{n,k} u_{n,k}(\rho) & = & 
\left( -\frac{\hbar^2\nabla^2_\rho}{2m} + \frac{\hbar^2 k^2}{2m} 
+ V_T(\rho) -\mu + \frac{2 gN}{L} \phi_0^2(\rho) \right) 
u_{n,k}(\rho) + \frac{gN}{L} \phi_0^2(\rho) v_{n,k}(\rho)
\label{eq:bog1}  \\
- \hbar \omega_{n,k} v_{n,k}(\rho) & = & 
\left( -\frac{\hbar^2\nabla^2_\rho}{2m} + \frac{\hbar^2 k^2}{2m} 
+ V_T(\rho) -\mu +  \frac{2 gN}{L} \phi_0^2(\rho) \right) 
v_{n,k}(\rho) + \frac{gN}{L} \phi_0^2(\rho) u_{n,k}(\rho)  \; .
\label{eq:bog2} 
\end{eqnarray}
\end{widetext}

Numerical solutions of the Bogoliubov equations for cylindrical 
condensates have already been obtained in the context of  theoretical 
studies of the Landau critical velocity \cite{fedichev}, the Landau 
damping of low energy collective oscillations \cite{muntsa}, and the 
behavior of solitary waves \cite{komineas}. Here we solve equations 
(\ref{eq:bog1})-(\ref{eq:bog2}) in order to use the functions $u$ and 
$v$ as a basis for a time dependent calculation, as explained in the 
next section. 

It is worth noticing that, expressing all lengths and energies in units 
of the harmonic oscillator length,  $a_\rho = [ \hbar / (m\omega_\rho 
)]^{1/2}$, and energy, $\hbar \omega_\rho$, respectively, the solutions 
of GP and Bogoliubov equations for the cylindrical condensate scale with 
a single parameter, $aN/L$. The same parameter can be expressed in terms 
of the chemical potential in the Thomas-Fermi limit, $\mu_{\rm TF}$. One 
has in fact,  $\eta \equiv  \mu_{\rm TF} / (\hbar \omega_\rho) = 
[ 4 a N/L]^{1/2}$ \cite{fedichev,note1}. We perform calculations for 
different values of $\eta$, which are representative of available 
elongated condensates. The lowest value that we consider is $\eta=9.4$, 
which simulates the condensate of Davidson and co-workers 
\cite{steinhauer2003,rehovot}, the largest value is $\eta=70$, for the 
condensate of Ref.~\cite{onofrio}. We also use an intermediate value, 
$\eta=26.5$, which corresponds to the condensate of the calculations 
in Refs.~\cite{brunello,brunello2}. 

\begin{figure}[htb]
\includegraphics[width=3.3in]{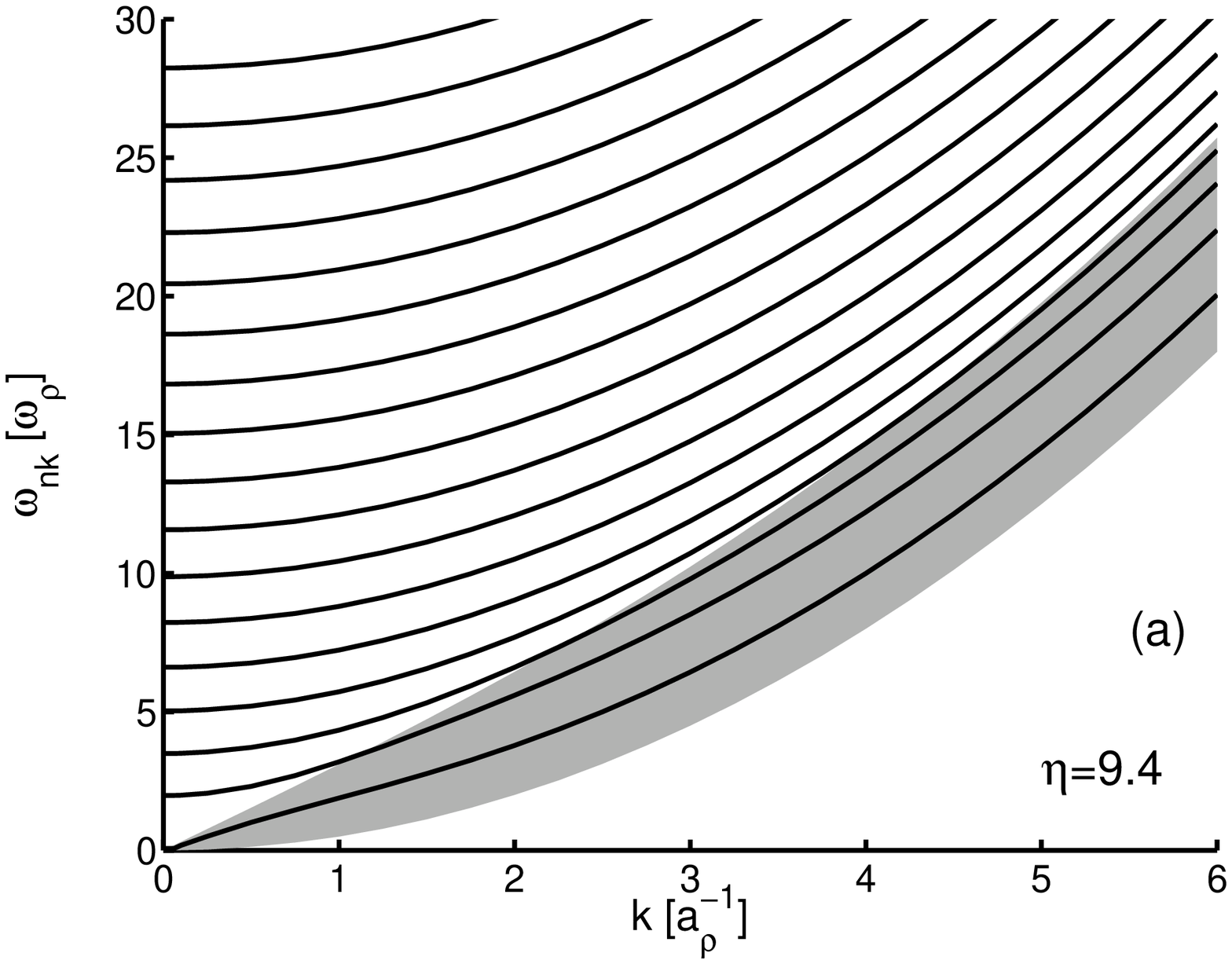}
\includegraphics[width=3.3in]{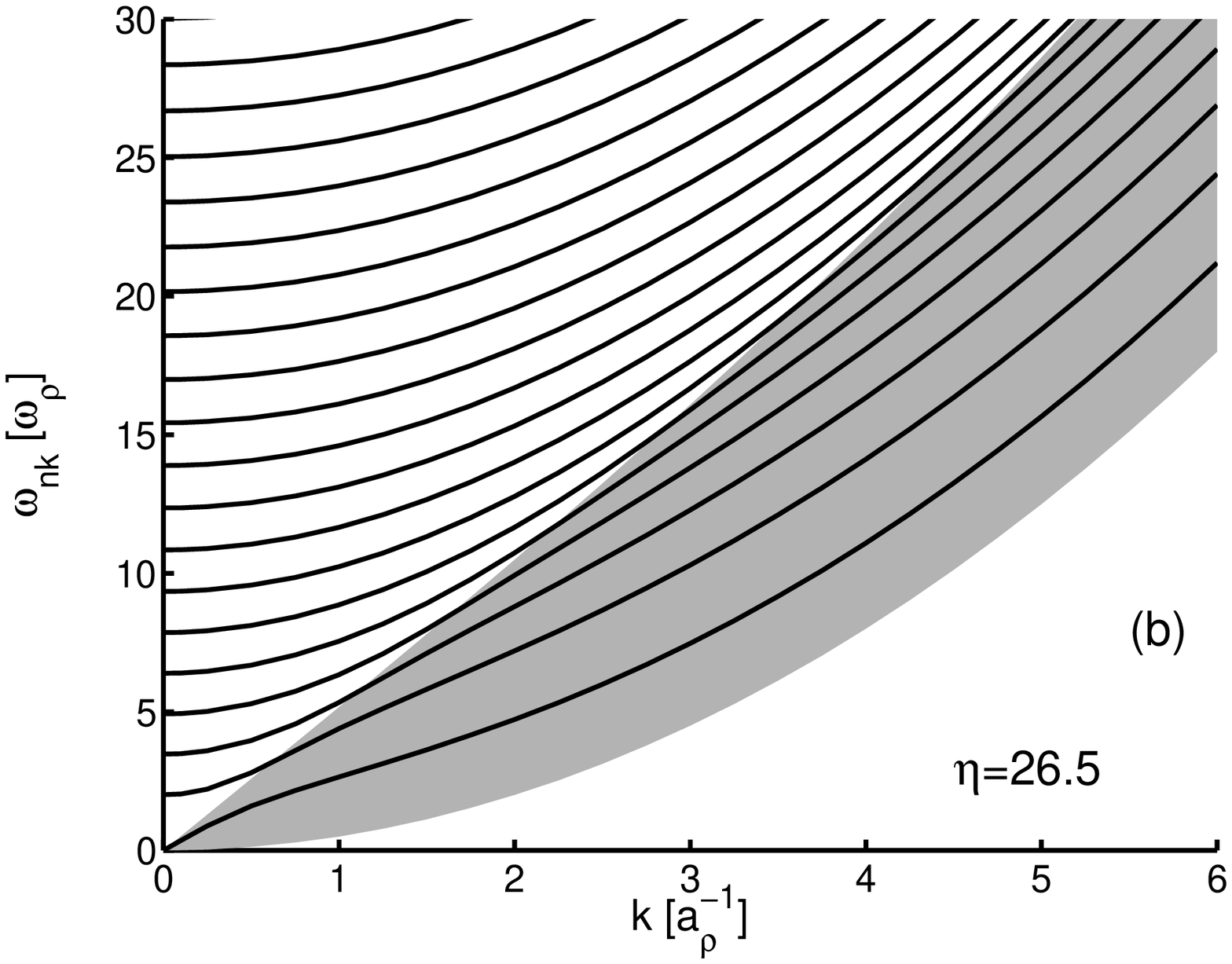}
\caption{ \label{fig:bogoliubov}
Spectrum of axially symmetric Bogoliubov excitations of cylindrical 
condensates with $\eta=9.4$ (a) and $\eta=26.5$ (b). The frequency 
$\omega_{n,k}$, in units of the radial trapping frequency $\omega_\rho$, 
is plotted as a function of the axial wavevector $k$, in units of 
$a_{\rho}^{-1}$, with $a_\rho= [\hbar/(m \omega_\rho)]^{1/2}$. The 
number of radial nodes is $n=0,1,2,\dots$, starting from the lowest 
branch. The shaded area corresponds to the region that gives finite 
contributions to the the dynamic structure factor $S(q,\omega)$ in 
local density approximation (see section~\protect\ref{sec:results1}).  
}
\end{figure}

In Figs.~1a and 1b we show the frequency $\omega_{n,k}$ of the 
Bogoliubov branches as a function of $k$, for $\eta=9.4$ and $26.5$, 
respectively. The lowest branch starts linearly at low $k$ and then 
becomes quadratic for large $k$, as expected for the textbook 
Bogoliubov spectrum of a uniform condensate. The transition between 
the two regimes occurs at $k$ of the order of $\xi^{-1}=\sqrt{\eta}$, 
where $\xi$ is the healing length. The slope of this branch 
has been the object of the investigation of Ref.~\cite{fedichev}. 
The second branch starts at $\omega_{2,0}= 2 \omega_\rho$, where 
it reduces to the transverse monopole (breathing) mode; its 
frequency is model independent as a consequence of a scaling 
property of GP equation in two dimensions \cite{pitaevskii97}. For 
$k \ll \xi^{-1}$ the dispersion law of the lowest branches can be 
obtained, with great accuracy, also within a hydrodynamic 
approach \cite{hydro}.  

Our purpose is now to explore the effects of the discrete 
multibranch spectrum of Figs.1a and 1b in the response of the 
condensate to a Bragg pulse.

\section{Time dependent quasiparticle amplitudes and momentum 
transferred}
\label{sec:bog(t)}

Let us suppose that a Bragg pulse is switched on at a time $t=0$. 
In the experiments, this is obtained by shining the condensate with
two laser beams with wavevectors ${\bf k}_1$ and ${\bf k}_2$ and 
frequencies $\omega_1$ and $\omega_2$. The action of the stimulated
light-scattering from atoms of the condensate is 
equivalent to a moving optical potential with wavevector ${\bf q} =
{\bf k}_1 - {\bf k}_2$ and frequency $\omega = \omega_1 - \omega_2$.  
Within the GP theory, this is accounted for by adding an extra term 
to the external potential, that becomes \cite{blakie}
\begin{equation}
V({\bf r},t)= V_T(\rho)  + \theta(t) V_B \cos(qz-\omega t) \, , 
\label{eq:vbragg}
\end{equation}
where $\theta(t)$ is the Heaviside step function and we have chosen 
${\bf q}$ along $z$.
We calculate the response of the condensate in two ways: i) by using
the quasiparticle projection method of Refs.~\cite{morgan,bbg}; ii) by
numerically solving the full time dependent GP equation (\ref{eq:TDGP})
for given values of $\eta, V_B, q$ and $\omega$. The latter case will
be discussed in the next section. Here we briefly sketch the 
quasiparticle projection method and its application to 
infinite cylindrical condensates. 

The starting point is a generalization of the Bogoliubov expansion 
(\ref{eq:linearized}) that allows the population of quasiparticle 
states to vary with time \cite{morgan}. The simplest way is to expand 
the order parameter in terms of a quasiparticle basis as in 
(\ref{eq:linearized}), but taking the coefficients $c_j$ to be
time dependent:
\begin{eqnarray}
\Psi({\bf r},t)& = &  e^{-i\mu t/\hbar} \{ \psi_0 ({\bf r}) + 
 \nonumber \\
 & \sum_j &  [ c_j (t) u_j({\bf r}) 
e^{-i \omega_j t} + c_j^*(t) v_j^*({\bf r})  
e^{i \omega_j t} ] \} \; .
\label{eq:t-exp}
\end{eqnarray}
The functions $\psi_0$, $u_j$ and $v_j$ are the solutions of 
Eqs.(\ref{eq:GP})-(\ref{eq:coupled2}) and are assumed to be static.    
Let us suppose that at $t=0$ the condensate is in the ground state 
and thus all $c_j(0)$ are zero. For $t>0$ the Bragg potential starts 
populating the quasiparticle states. The dynamics is entirely fixed by 
the evolution of the coefficients $c_j(t)$, that can be obtained by 
inserting the expansion (\ref{eq:t-exp}) into the GP equation 
(\ref{eq:TDGP}), with the external potential given in (\ref{eq:vbragg}). 
One finds \cite{bbg}
\begin{eqnarray}
\sum_j & [ & \dot{c}_j (t) u_j({\bf r}) e^{-i \omega_j t} 
+ \dot{c}_j^*(t) v_j^*({\bf r}) e^{i \omega_j t} ] \nonumber \\
& = & \frac{\theta(t) V_B}{i \hbar}\ e^{-i\mu t /\hbar} 
\cos(qz-\omega t) \Psi({\bf r},t) \; . 
\end{eqnarray}
In order to get an expression for the $\dot{c}$'s, one can 
multiply by $u_i^*$ and $v_i^*$ the last equation and its conjugate,
respectively, and integrate in $d{\bf r}$. By using the orthogonality
and symmetry relations (\ref{eq:ortho}) one gets   
\begin{eqnarray}
\dot{c}_i(t) = \frac{\theta(t) V_B}{i \hbar}\ e^{i \omega_i t} 
\int \! d{\bf r} & [ & u_i^* e^{i \mu t} \Psi  +  
v_i^* e^{-i \mu t} \Psi^*  ] \nonumber \\ 
& \times & \cos(qz -\omega t)
\end{eqnarray}
Finally, assuming that the quasiparticle occupation is always much 
smaller than $N$, one can replace $\Psi$ with the ground state order
parameter and integrate in time. The results is \cite{bbg}
\begin{equation}
c_i(t) = \frac{V_B}{i \hbar} \int_0^t \! dt^\prime  
e^{i \omega_i t^\prime} 
\int \! d{\bf r} \ 
(u_i^* + v_i^*) \psi_0 \cos (qz - \omega t^\prime) 
\end{equation}
which is expected to be valid in the limit of small perturbations. 

In the case of a cylindrical condensate, we can use the factorization 
(\ref{eq:uv-cylinder}) to rewrite the previous expression in the
form 
\begin{eqnarray}
c_{n,k}(t) =  \frac{ V_B 2 \pi \sqrt{N} }{i \hbar L} \int_0^t dt^\prime \ 
e^{i \omega_{n,k} t^\prime} \int_0^\infty \! d\rho \ \rho [u^*_{n,k}(\rho) 
\nonumber \\ 
 + v^*_{n,k} (\rho)] \phi_0(\rho) \int_{-\infty}^\infty \! dz \ e^{-ikz} 
\cos(qz-\omega t^\prime) \; . 
\end{eqnarray}
The last integral simply selects out the quasiparticles with 
$k=\pm q$, so that 
\begin{eqnarray}
c_{n,k}(t) =  - \frac{ \sqrt{N} V_B W_{n,k}}{2 \hbar}  \left[  
\frac{ e^{i(\omega_{n,k} - \omega)t} -1}{\omega_{n,k} - \omega} 
\delta_{k,q} \right. \nonumber \\ 
\left. +  \frac{ e^{i(\omega_{n,k} + \omega)t} -1}{\omega_{n,k} 
+ \omega} \delta_{k,-q}
\right] \; , 
\label{eq:cnkt}
\end{eqnarray}
where 
\begin{equation}
W_{n,k} = 2 \pi \int \! d\rho \ \rho [u^*_{n,k}(\rho) 
+ v^*_{n,k}(\rho)] \phi_0(\rho) \; . 
\label{eq:wnk}
\end{equation}
This means that, once the ground state and the Bogoliubov spectrum 
are known at $t=0$, the order parameter at time $t>0$ can be calculated 
in a rather simple way. 

An important quantity that can easily be calculated within this
scheme is the total momentum imparted to the condensate, which can 
be defined by integrating the current density associated with the 
order parameter: 
\begin{equation}
P_z(t) = \frac{\hbar }{2i} \int \! d{\bf r}\ \Psi^* ({\bf r},t) 
\frac{\partial }{\partial z} \Psi ({\bf r},t)  + {\rm c.c.} \; . 
\label{eq:pzdefinition}
\end{equation}
By inserting the expansion (\ref{eq:t-exp}) and using the orthogonality
and symmetry relations (\ref{eq:ortho}), one gets the simple expression
\begin{equation}
P_z (t) = \sum_{n,k} \hbar k \ | c_{n,k}(t) |^2 \; . 
\end{equation}
Now, let us take $c_{n,k}(t)$ from Eq.~(\ref{eq:cnkt}) and use the 
fact that the eigenfrequencies and eigenfunctions of the Bogoliubov 
equations (\ref{eq:bog1})-(\ref{eq:bog2}) are even functions of $k$.
One gets the final result
\begin{eqnarray}
P_z (t)= \frac{ N q V_B^2 t^2 }{4 \hbar} \sum_n |W_{n,q}|^2 
\left[ \left( 
\frac{\sin[(\omega_{n,q}-\omega)t/2]}{(\omega_{n,q}-\omega)t/2} 
\right)^2 \right.  \nonumber \\
\left. -  \left(
\frac{\sin[(\omega_{n,q}+\omega)t/2]}{(\omega_{n,q}+\omega)t/2} 
\right)^2 \right] .
\label{eq:finalpz} 
\end{eqnarray}
In the next section we will show typical results obtained with 
this expression, but some relevant features are already evident. 
Notice first that, for positive $\omega$ and a given $q$, the 
momentum transferred (\ref{eq:finalpz}) is clearly resonant 
at the frequencies $\omega=\omega_{n,q}$. In other words, a peak 
occurs in $P_z$ whenever a vertical line at $k=q$ crosses the
Bogoliubov branches of Figs.~1a and 1b. The separation between 
these branches is roughly $1$-$2\ \omega_\rho$, while the width 
of each peak goes like $2 \pi/t$. Thus, in order to resolve the 
different peaks, one has to wait at least a time of the order of 
trapping period $T_\rho = 2\pi /\omega_\rho$. For $t$ in this 
range, the second term in the square bracket gives certainly a 
negligible contribution. Finally, the contribution of 
each axial branch to the total momentum depends on the quantities 
$W_{n,q}$, defined in Eq.~(\ref{eq:wnk}). Typically, for a given 
$q$, the lowest branches have a greater weight in the summation 
(\ref{eq:finalpz}), since the corresponding quasiparticle
amplitudes have a greater overlap with the ground state.  

This behavior of $P_z$ reflects its direct connection with the 
dynamic  structure factor. At zero temperature, the latter is 
defined as
\begin{equation}
S({\bf q},\omega) = \sum_i |\langle i | \hat{\rho}_{\bf q} 
| 0 \rangle |^2 \delta (\omega - \omega_i) \; , 
\end{equation}
where $\hat{\rho}_{\bf q}$ is the density fluctuation operator, 
$|0\rangle$ is the ground state, $|i \rangle$ are the excited states, 
and $\omega_i = (E_i - E_0)/\hbar$. In terms of the quasiparticle 
amplitudes one has \cite{csordas,zambelli}
\begin{equation}
\langle i | \hat{\rho}_{\bf q} | 0 \rangle = 
\int \! d{\bf r} \ [ u^*_i ({\bf r}) + v^*_i ({\bf r})] 
e^{i {\bf q} \cdot {\bf r} } \psi_0({\bf r}) \; . 
\end{equation}
If ${\bf q}$ is taken along $z$, and the system is translational 
invariant in the same direction, one can use Eqs.~(\ref{eq:uv-cylinder}) 
and (\ref{eq:wnk}) to rewrite $S({\bf q},\omega)$ as
\begin{equation}
S(q,\omega)= \sum_{n} N |W_{n,q}|^2 \delta(\omega-\omega_{n,q}) \,.
\label{eq:sqomega}
\end{equation}
Comparing this result with Eq.~(\ref{eq:finalpz}) one finds 
\cite{bbg,brunello}
\begin{equation}
\lim_{t\to \infty} P_z (t) = \frac{ \pi q V_B^2 t }{2 \hbar}
[ S(q,\omega) - S(-q,-\omega)] \; . 
\end{equation}
As discussed in Ref.~\cite{bbg}, these two quantities are connected 
in such a simple way only for a cylindrical condensate. If the condensate 
is trapped also along $z$, the relation between  $S(q,\omega)$ and 
$P_z$ is less direct, involving a further integration on the 
duration time of the Bragg pulse. However, if the axial trapping
frequency is significantly smaller than the radial one, there exists 
a wide range of time where the connection between the momentum
transferred and the dynamic structure factor is roughly the same
as for an infinite cylinder. In that range, the Bogoliubov branches,
which correspond to $\delta$ peaks in $S(q,\omega)$, are also 
observable as distinguishable peaks in $P_z(t)$, as we will see in
section V. 

\begin{figure}
\includegraphics[width=3.3in]{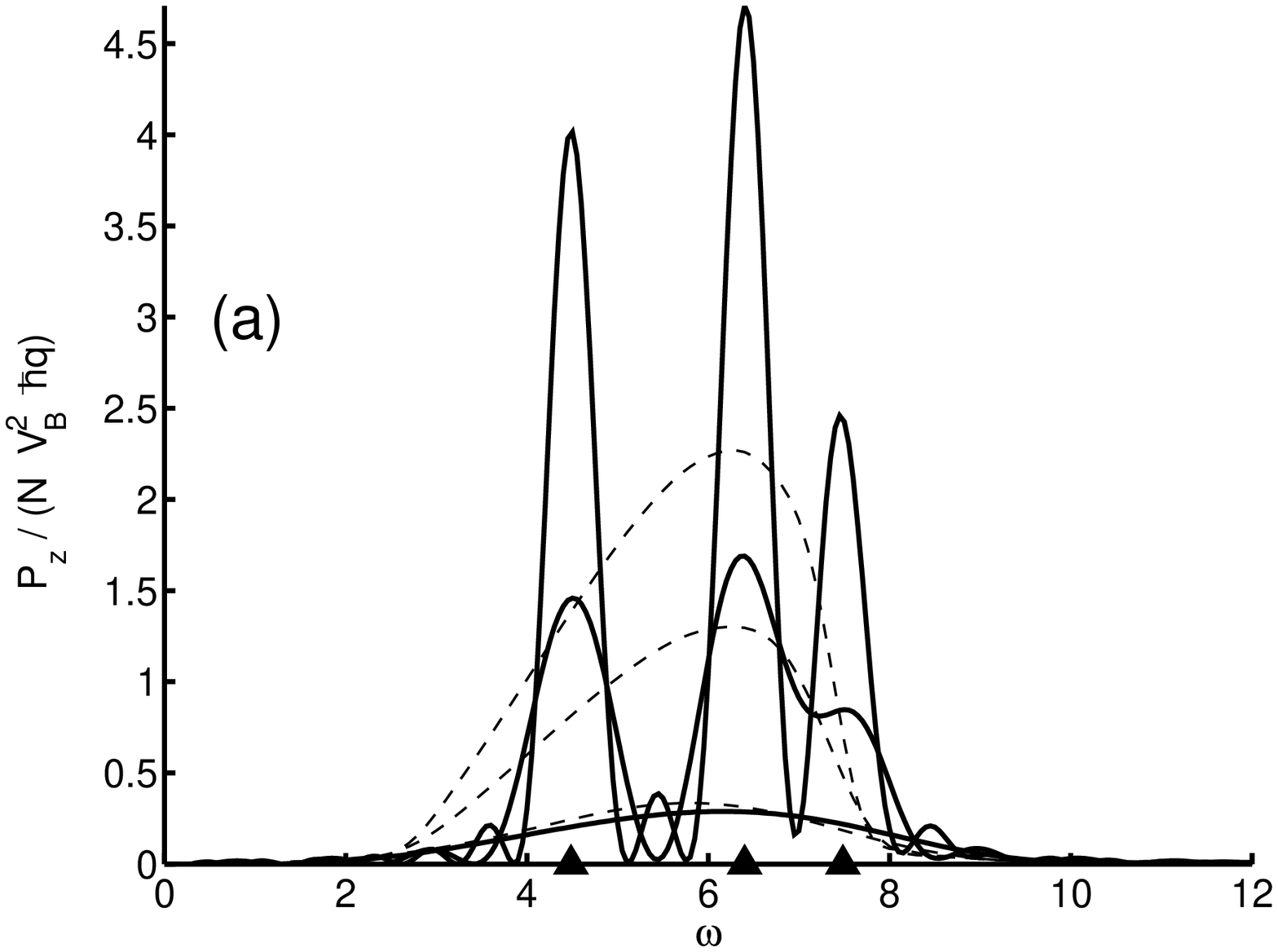}
\includegraphics[width=3.3in]{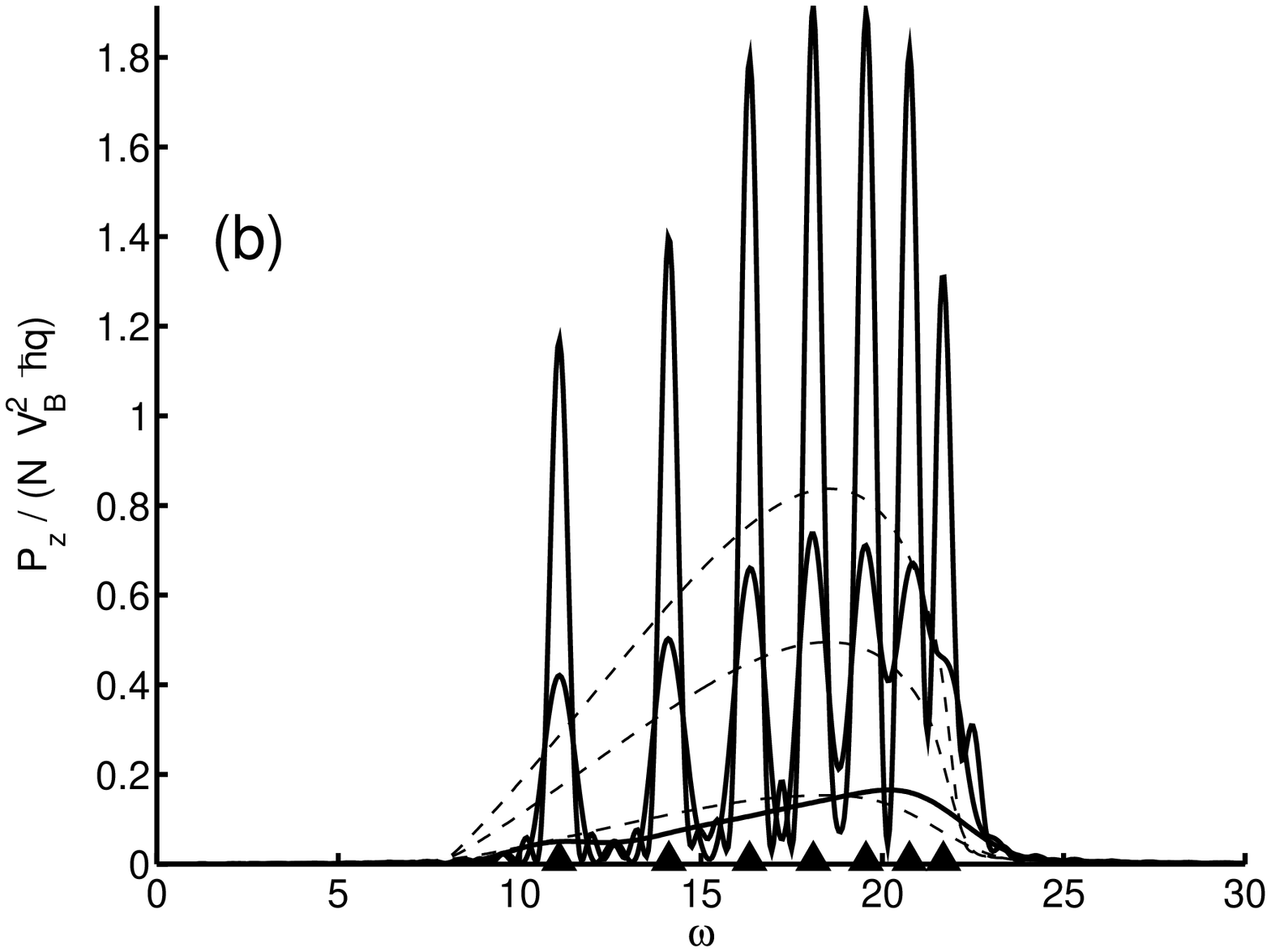}
\caption{ \label{fig:pz-bog}
Momentum transferred to cylindrical condensates with $\eta=9.4$ (a)
and $\eta=26.5$ (b) by a Bragg pulse of wavevector $q=2.3 a_\rho^{-1}$ 
(a) and $q=4\ a_\rho^{-1}$ (b), and frequency $\omega$. The quantity 
$P_z/(N V_B^2 \hbar q)$, in units of $(\hbar \omega_\rho)^{-2}$, is 
plotted vs. $\omega$, in units of $\omega_\rho$. Solid lines 
correspond to the results obtained from Eq.~(\protect\ref{eq:finalpz}) 
at three different times, $t=2$ (bottom curve), $6$ (mid) and $10$
(top)  $\omega_\rho^{-1}$. The dashed lines are the corresponding 
predictions of the local density approximation. The triangles on the 
horizontal axis are the frequencies of the lowest Bogoliubov 
states that contribute to $P_z$.  
}
\end{figure}

\begin{figure}
\includegraphics[width=3.3in]{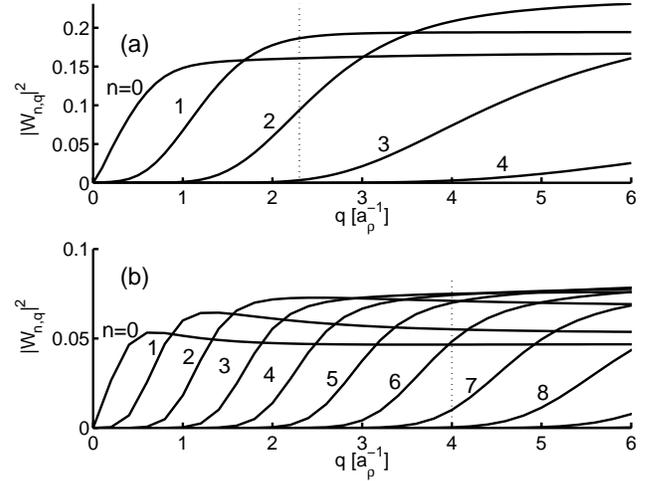}
\caption{ \label{fig:wnq}
Quantities $|W_{n,q}|^2$, defined in Eq.(\protect\ref{eq:wnk}), for the
two condensates in Fig.~2. In the linear response regime, these
functions provide the height of the $\delta$-peaks in the dynamic
structure factor (\protect\ref{eq:sqomega}), as well as of the peaks in 
the momentum transferred (\protect\ref{eq:finalpz}). The
values of $q$ used in Fig.~2 are shown as vertical dotted lines.  
}
\end{figure}

\section{Bogoliubov branches and multipeak spectrum}  
\label{sec:results1}

In Figs.~2a and 2b we show the momentum transferred to two cylindrical 
condensates with $\eta=9.4$ and $26.5$ by a Bragg pulse with wavevector  
$q=2.3 a_\rho^{-1}$ (a) and $q=4\ a_\rho^{-1}$ (b). Recalling that 
$\xi^{-1} = \sqrt{\eta}$, the values of $q$ are chosen in order to 
have roughly the same value of $q\xi$ in the two cases. As solid lines
we plot the quantity $P_z(t)/(N V_B^2 \hbar q)$ as a function of 
$\omega$ and for three different durations times: $t=2, 6$ and $10\ 
\omega_\rho^{-1}$. According to Eq.~(\ref{eq:finalpz}), 
this quantity is independent of $V_B$.  

For a pulse shorter than the trapping period, the response is 
distributed over a single broad peak. Short pulses were used in the 
first experiments of Refs.~\cite{mit,rehovot}. In this regime, 
accurate predictions can be found by using the local density 
approximation (LDA) \cite{zambelli,brunello}, i.e., assuming 
the system as locally uniform. If the excitation spectrum is the 
one of a uniform gas, but with a dispersion law fixed by the 
local density $n({\bf r})$, the dynamic structure factor $S(q,\omega)$ 
turns out to be non zero only within the frequency range, $\omega_r 
< \omega < \omega_r [1 + 2 \mu/(\hbar \omega_r)]^{1/2}$, where $\hbar 
\omega_r = \hbar^2 q^2 /(2m)$ is the free recoil energy \cite{zambelli}. 
This range corresponds to the shaded area in Figs.~1a and 1b. The 
LDA approximation for $S(q,\omega)$ can be used to estimate the 
momentum transferred $P_z(t)$ as in \cite{brunello}. The corresponding 
curves are shown as dashed lines in Figs.~2a and 2b. The agreement 
between dashed and the solid lines for lowest curve at $t=2 
\omega_\rho^{-1}$ is reasonable; for shorter times we reproduce 
the results of Ref.~\cite{brunello}, and the agreement with LDA 
becomes better and better. 

For $t$ of the order of $T_\rho = 2 \pi \omega_\rho^{-1}$ new 
structures begin to appear, strongly deviating from LDA. In particular, 
one finds a multipeak spectrum that reflects the existence of 
Bogoliubov branches, i.e, the same plotted in Fig.~1. The peak at 
the lowest frequency is the collective axial mode with no nodes 
in the radial direction ($n=0$). The $n=1$ and $n=2$ branches are also 
visible for the condensate with $\eta=9.4$ in (a), and even more 
for $\eta=26.5$ in (b). The corresponding frequencies, $\omega_{n,q}$, 
are indicated by triangles on the horizontal axis. 

From Eq.~(\ref{eq:finalpz}) one sees that the height of each peak, 
for $t \gg T_\rho$, becomes proportional to $t^2 |W_{n,q}|^2$. 
In Figs.~3 we plot $|W_{n,q}|^2$ as a function of $q$ for the same 
condensates of Fig.~2. This figure tells us how many branches
give significant contributions to $P_z$ and $S(q,\omega)$ at a 
given $q$. One notices that the contribution of each branch is
sizeable when the corresponding $\omega_{n,q}$ is, roughly speaking, 
within the shaded area in Fig.1. The first branch is the dominant
one at small $q$. Its weight $|W_{0,q}|^2$ also vanishes at $q \to 0$, 
in agreement with the limiting behavior of long wavelength phonons, 
for which $u_{0,q} \simeq - v_{0,q} \propto q^{-1/2}$, so that the
integral (\ref{eq:wnk}) vanishes. One also notices that, for a given
$q$, the number of branches that contribute to the summation 
(\ref{eq:finalpz}) increases with $\eta$. This effect is rather 
dramatic for the condensates used by Ketterle and co-workers
\cite{mit}, where $\eta \gg 10$ (see Fig.~4). In this case, even a 
small extra broadening in the experimental detection of $P_z$, 
might significantly reduce the visibility of the Bogoliubov 
branches.  
       
\begin{figure}[htb]
\includegraphics[width=3.3in]{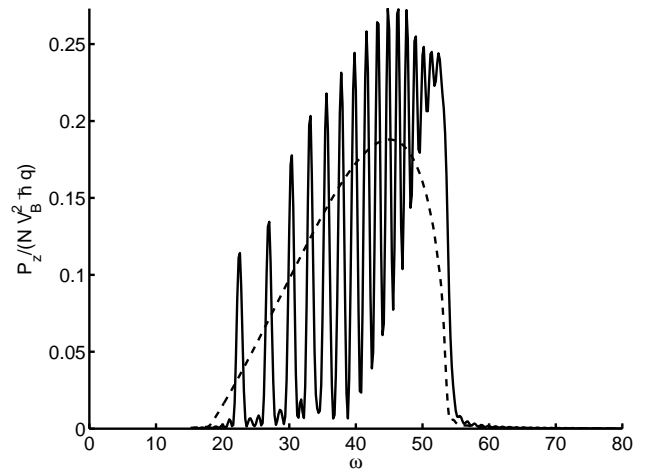}
\caption{ \label{fig:pz-mit}
Same as in Fig.~2, but with $\eta=70$, $q=6 \ a_\rho^{-1}$ and  
$t=6 \ \omega_\rho^{-1}$. 
}
\end{figure}

The momentum transferred can also be calculated by direct 
integration of the time-dependent Gross-Pitaesvkii equation 
(\ref{eq:TDGP}), using the definition (\ref{eq:pzdefinition}) 
for $P_z$. In the linear response regime the 
results must concide with the ones of the quasiparticle projection 
method so far discussed. An example is shown in  Fig.~5. The solid 
line corresponds to the momentum calculated with Eq.~(\ref{eq:finalpz}) 
at $t=6 \omega_\rho^{-1}$, as in Fig.~2a. The empty squares are the 
results of GP simulations for a weak Bragg pulse ($V_B=10^{-3} \ 
\hbar \omega_\rho$). As expected, the two results 
are in good agreement, except for small differences that are compatible 
with the accuracy of our GP simulations. We also checked that, for 
such small values of $V_B$, the quantity $P_z(t)$ scales with 
$V_B^2$ as expected for the linear response. 
 
A major advantage of using both approaches, is that one can now 
increase $V_B$, beyond the linear regime, and look at the differences 
between the linear response, given by Eq.~(\ref{eq:finalpz}), and the 
numerical GP calculation, which is valid also in the nonlinear regime 
\cite{bbg,band}. A typical result is shown in Fig.~5, where we compare 
the small $V_B$ limit (solid line and empty squares) with the
results of GP simulations for $V_B=0.5 \ \hbar \omega_\rho$ (solid
circles). With such a pulse intensity, the fraction of condensate 
atoms which are excited is roughly $25\%$ at resonance. This value
is slightly larger than that of typical experiments, where a fraction 
of the order of $10$-$20\%$ is required for the optical detection of the 
excited atoms. The figure shows that, even for such highly excited 
condensates, the main features of the multipeak response, associated
with the underlying Bogoliubov spectrum, are still visible. The most 
significant effects of nonlinearity, originating from the mean-field 
interaction in the GP equation, are that i) $P_z$ does not scale
with $V_B^2$,  ii) the peak frequencies are slightly shifted downwards,
and iii) additional structures appear superimposed to the original 
shape of the peaks. This behavior is a consequence of typical nonlinear
processes like mode-mixing and higher harmonic generation.  Similar 
effects were already predicted for spherical \cite{bbg} and one-dimensional 
\cite{band} condensates. 

\begin{figure}[htb]
\includegraphics[width=3.3in]{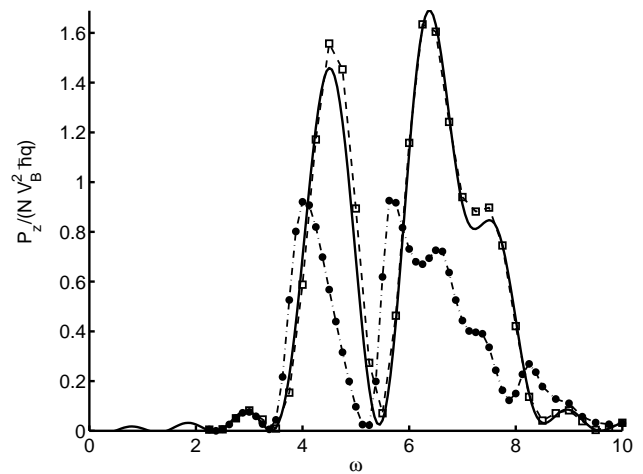}
\caption{ \label{fig:pz-gp}
Momentum transferred to a cylindrical condensate with $\eta=9.4$ by 
a Bragg pulse with $q=2.3\ a_\rho^{-1}$. The quantity $P_z/(N V_B^2 
\hbar q)$, in units of $(\hbar \omega_\rho)^{-2}$, is plotted vs. 
$\omega$, in units of $\omega_\rho$. The solid line corresponds 
to the result obtained 
from Eq.~(\protect\ref{eq:finalpz}) at $t=6 \ \omega_\rho^{-1}$
(same as in Fig.~2a). Empty squares and solid circles 
are the results of the numerical integration of the time dependent 
GP equation (\protect\ref{eq:TDGP}) for a Bragg pulse of intensity 
$V_B=10^{-3} \ \hbar \omega_\rho$ and $V_B=0.5 \ \hbar \omega_\rho$, 
respectively. Dashed and dot-dashed lines are a guide to the eye.  
}
\end{figure}

\section{Bragg scattering from a prolate elongated condensate}
\label{sec:results2}

Here we show that the main features found for the response of
a cylindrical condensate to a Bragg pulse remain observable also
in a finite elongated condensate. 

Let us take the trapping potential in the form
\begin{equation}
V_T({\bf r}) = (1/2) m \omega_\rho^2 (\rho^2 + \lambda^2 z^2)
\label{eq:trap}
\end{equation}
where $\lambda=\omega_z/\omega_\rho$. If $\lambda < 1$ the
condensate at equilibrium is a prolate ellipsoid. The ground 
state at $t=0$ can be found as the stationary solution
of Eq.~(\ref{eq:TDGP}).  Then, the time dependent GP equation 
can be solved at $t>0$ to simulate the Bragg process. 

In Figs.~6a and 6b we show typical results obtained for $P_z$
in the linear response (small $V_B$) regime. We simulate the 
Bragg scattering from two different condensates, whose chemical 
potential, in units of $\hbar \omega_\rho$, is equal to $9.4$ and 
$26.5$, as for the cylindrical condensates of Fig.~1 and 2. These 
two values correspond to the condensate used in the experiments 
of Ref.~\cite{steinhauer2003}, made of $N=10^5$ atoms of 
$^{87}$Rb in a trap with $\omega_\perp= 2\pi (220$Hz$)$ and 
$\lambda=0.114$, and to the one used in the calculations of 
Refs.~\cite{brunello,brunello2}, with $\lambda=0.125$, 
respectively. With this choice, the condensates in Fig.~2a 
(2b) and 6a (6b) have also the same density profile in the 
radial direction for $z=0$. As one can see, the momentum 
transferred to the trapped condensates of Fig.~6 has the 
same multipeak structure of the corresponding cylindrical 
condensates of Fig.~2.  

\begin{figure}[htb]
\begin{center}
\includegraphics[width=3.3in]{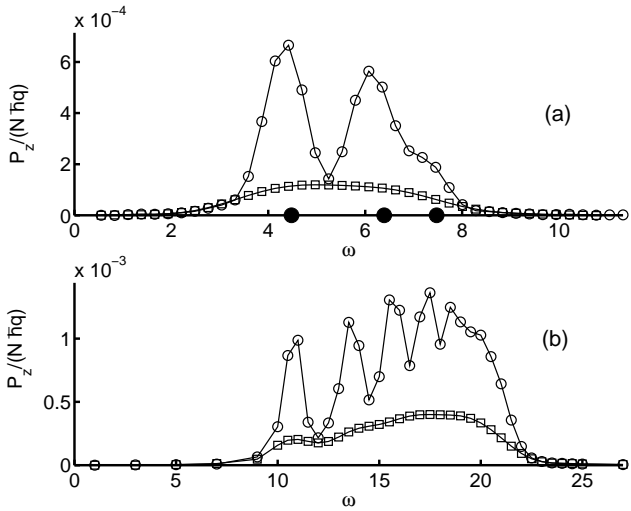}
\end{center}
\caption{ \label{fig:pz-trapped}
Momentum transferred to two different ellipsoidal condensates 
by a Bragg pulse. In part (a) the condensate has $\mu = 9.4 \hbar
\omega_\rho$ and $\lambda=0.114$; the Bragg pulse has $q=2.3 
a_\rho^{-1}$ and $V_B=0.02 \hbar \omega_\rho$. In part (b) the
condensate has $\mu = 26.5 \hbar \omega_\rho$ and $\lambda=0.125$; 
the Bragg pulse has $q= 4 a_\rho^{-1}$ and $V_B=0.05 \hbar 
\omega_\rho$. The points are obtained by numerically integrating 
the GP equation (\protect\ref{eq:TDGP}) for different $\omega$
(in units of $\omega_\rho$). The duration 
time is $t=2$ and $6\ \omega_\rho^{-1}$ for the lower and upper 
curves, respectively. The lines are a guide to the eye. The 
solid circles on the horizontal axis in (a) are the frequencies 
of the lowest Bogoliubov branches for that $q$.  }
\end{figure}
  
The response of an ellipsoidal (axially confined) condensate 
is expected to differ from the one of a cylindrical (axially 
unbound) condensate for two main reasons. First, their Bogoliubov
spectra are different: the excitations of a confined condensate
are discretized also along $z$ and the quasiparticle amplitudes
$u$ and $v$ have a nontrivial dependence on $\rho$ and $z$. 
Second, the response for long pulses is affected by the presence
of the axial confinement, which can produce a reflection of
quasiparticles at the boundaries and a center-of-mass 
motion of the condensate in the trap. 

The first effect, however, is small if $q \gg L^{-1}$, where
$L$ is the axial length of the condensate. In this case, the Bragg
pulse excites quasiparticles that behave like travelling waves
along $z$, having wavelength much shorter than the condensate
size. One can safely introduce a continuous wavevector $k$ to
identify such excitations as in the multibranch spectrum of
the cylindrical condensate. The axial size of the two condensates 
in Figs.~6a and 6b is approximately $L=75$ and $120 a_\rho$, 
respectively, so that the Bogoliubov modes can be safely 
represented by continuous branches down to about $k=0.5 \ a_\rho^{-1}$. 
Moreover, for $k \gg L^{-1}$, the frequencies $\omega_{n,k}$ for 
the elongated ellipsoidal condensate are expected to be close to the 
ones of the corresponding cylindrical condensate, with a similar 
multibranch structure. In order to check that the peaks of $P_z$ 
are still located at the frequencies of the Bogoliubov branches, 
we make a further simulation for the condensate in Fig.~6a. 
We excite the condensate by acting with a Bragg pulse for a 
given duration $t$ and then we let it to freely evolve in the 
trap. We perform a Fourier analysis of the induced density variations 
in the axial direction. As expected, we find that, for each $q$ 
and $\omega$, the density oscillates as a superposition of waves 
of different frequencies $\omega_{n,k}$. The latter are extracted 
from the Fourier spectrum and reported in Fig.~6a as solid circles. 
As one can see, the momentum is resonantly transferred at 
those frequencies.

The second effect is also small if the duration time of the Bragg 
pulse is significantly less than the axial trapping period $T_z$.
If $\lambda$ is of the order of $10^{-1}$, or less, there 
exists a sufficiently wide interval of time, between $T_\rho$ 
and $T_z$, where the multipeak structures of $P_z$ can be
resolved without being significantly affected by the reflection
of quasiparticles at the boundaries and by the center-of-mass
motion. 

\begin{figure}[htb]
\begin{center}
\includegraphics[width=3.3in]{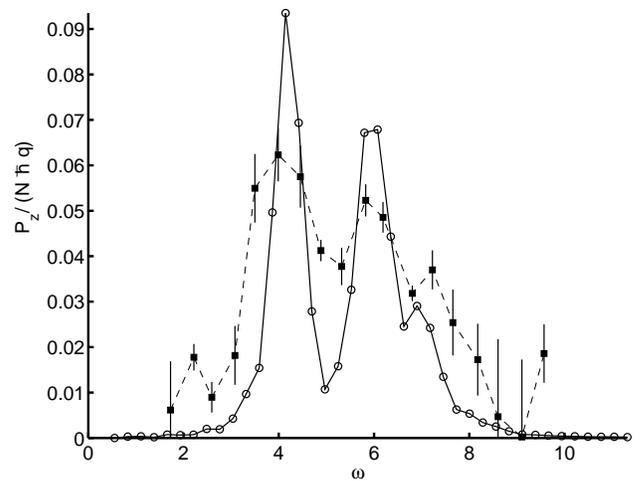}
\end{center}
\caption{ \label{fig:pz-davidson}
Momentum transferred to the condensates of 
Ref.~\protect\cite{steinhauer2003}. Empty symbols are obtained 
by numerically  integrating the GP equation (\protect\ref{eq:TDGP}) 
for different values of $\omega$ (in units of $\omega_\rho$). 
Points with error bars are experimental data 
\protect\cite{steinhauer2003}. The lines are a guide to the 
eye.  The condensate is the same of Fig.~6a. The Bragg pulse has
wavevector $q=2.3 a_\rho^{-1}$, duration $t= 6$ ms$=8.29 
\omega_\rho^{-1}$ and intensity $V_B=0.2 \hbar \omega_\rho$. 
}
\end{figure}

Finally, the experimental observation of the excited atoms, 
requires rather high Bragg intensities. As already shown in the 
previous section, nonlinear processes are not expected to 
hinder the visibility of the multibranch Bogoliubov spectrum. 
Conversely, the combined use of both the quasiparticle projection
method and the time dependent GP equation might allows one
to extract from the observed $P_z$ information about nonlinear
effects in the condensate dynamics. 
  
In Fig.~7 the results of our GP simulations are compared with 
experimental measurements of $P_z$. Similar examples were already 
reported in Ref.~\cite{steinhauer2003}, where the major sources 
of noise and broadening of the observed data were also discussed. 
The main message of Ref.~\cite{steinhauer2003} was that the measured 
response to long Bragg pulses exhibits clear evidences of the 
underlying Bogoliubov spectrum. In the present paper we have provided
a deeper theoretical analysis, which supports this view.

\begin{acknowledgments}
This research was supported by MIUR, Project PRIN-2002. We are 
indebted to N. Davidson, J. Steinhauer, N.Katz and R. Ozeri for 
many useful discussions. F.D. likes to thank the Dipartimento di 
Fisica of the Universit\`a di Trento for hospitality.   
\end{acknowledgments}

\appendix

\section*{NUMERICAL PROCEDURES 
\label{appendix} }

\begin{itemize}

\item{ i) Stationary GP equation } 

The ground state configuration at $t=0$ is obtained by solving the
stationary GP equation (\ref{eq:GP}) with the external potential
given in (\ref{eq:trap}). The case $\lambda=0$ corresponds to the
cylindrical condensate of sections~III and IV:  the
order parameter is uniform along $z$ and the problem is reduced
to the one-dimensional equation (\ref{eq:phi0}). The solution of
this equation is equivalent to the minimization of a GP energy 
functional \cite{rmp}. We find the minimum by mapping  the 
function $\phi_0$ on a grid of points and propagating it 
in imaginary time with an explicit first order algorithm, starting
from a trial configuration, as described in Ref.~\cite{ds}.
The case $\lambda \neq 0$ corresponds to an ellipsoidal 
condensate, as in section V.  We use the same minimization 
algorithm. Due to the axial symmetry of $V_T$, the problem is 
two-dimensional. The order parameter is mapped on a  
$N_\rho \times N_z$ grid (typically, $64 \times 1024$ points). 

\item{ ii) Time dependent GP equation }

Let us define an effective Hamiltonian $H$ by rewriting 
Eq.~(\ref{eq:TDGP}) in the form $i \hbar \partial_t \Psi  = H \Psi$.
We solve this equation by propagating the order parameter 
in real time, starting from the ground state configuration.
At each time step, $\Delta t$, the evolution is determined by 
the implicit algorithm  
\begin{equation}
\left( 1+ \frac{i \Delta t}{2} H \right) \Psi_{n+1} =
\left( 1- \frac{i \Delta t}{2} H \right) \Psi_n
\label{eq:steps}
\end{equation}
where $H$ and $\Psi_n$ are calculated at the $n$-th iteration. For
$\lambda \neq 0$, the propagation is splitted into the axial and 
radial parts.  The former is obtained by using a fast Fourier
transform algorithm to treat 
the kinetic energy term, while the latter is  performed with a
Crank-Nicholson differencing method, as in \cite{brunello}.
For the cylindrical case ($\lambda=0$), we first write the
order parameter as $\Psi(\rho,z,t) = \sum_\nu \psi_\nu(\rho,t)
\exp(i\nu qz)$, where $q$ is the wavevector of the Bragg potential,
and then we propagate in time the radial functions $\phi_\nu$,
for $\nu=0,\pm 1,\pm 2,\dots$. Only a few values of $\nu$ give
significant contributions to the sum, their number depending 
on the intensity of the pulse.  

\item{ iii) Bogoliubov equations }

In order to solve the eigenvalue problem (\ref{eq:bog1})-(\ref{eq:bog2}) 
for a cylindrical condensate, we project the unknown radial 
functions $u(\rho)$ and $v(\rho)$ on a basis set of Bessel functions. The
problem is thus reduced to a matrix diagonalization.  We use Bessel 
functions with up to 100 nodes in the computational box. The radius 
of the box is typically two or three times the size of the condensate.  

\end{itemize}


\begin{thebibliography}{24}
\bibitem{note1} In this context, the Thomas-Fermi approximation 
                corresponds to neglecting the quantum pressure term, 
                $\nabla^2 \phi_0$, in the stationary GP equation 
                (\protect\ref{eq:phi0}). The density $\phi_0^2(\rho)$
                takes the form of an inverted parabola and the
                chemical potential can be easily calculated from 
                the normalization.
\end{thebibliography}
\end{document}